\documentclass[%
 aip,
 reprint,
 rsi,
 amsmath,amssymb,
]{revtex4-2}

\usepackage{graphicx}
\usepackage{dcolumn}
\usepackage{bm}
\usepackage{mathrsfs}

\usepackage[version=4]{mhchem}

\usepackage[caption=false]{subfig}

\newcommand{\ket}[1]{\vert #1 \rangle}

\newcommand{\bra}[1]{\langle #1 \vert}

\newcommand{\Dbraket}[2]{\langle #1 \hspace{.10em} \vert \hspace{.10em}  #2 \rangle}
\newcommand{\Tbraket}[3]{\langle #1 \hspace{.10em} \vert \hspace{.10em} #2 \hspace{.10em} \vert \hspace{.10em} #3 \rangle}

\newcommand{\DBraket}[2]{\Big\langle #1 \hspace{.10em} \Big\vert \hspace{.10em}  #2 \Big\rangle}
\newcommand{\mbf}[1]{\boldsymbol{#1}}

\newcommand{\hf}{\mathrm{HF}}

\renewcommand{\phi}{\varphi}

\begin{document}

\preprint{APS/123-QED}

\title{Communication: Non-adiabatic derivative coupling elements for the coupled cluster singles and doubles model}

\author{Eirik F.~Kj{\o}nstad}
\email{eirik.kjonstad@ntnu.no}
\affiliation{Department of Chemistry and The PULSE Institute, Stanford University, Stanford, California 94305, USA}
\affiliation{ 
Department of Chemistry, Norwegian University of Science and Technology, 7491 Trondheim, Norway
}%
\author{Henrik Koch}
\affiliation{ 
Department of Chemistry, Norwegian University of Science and Technology, 7491 Trondheim, Norway
}
\affiliation{%
Scuola Normale Superiore, Piazza dei Cavaleri 7, 56126 Pisa, Italy 
}%
%


\date{\today}

\begin{abstract}
We present an efficient implementation of analytical non-adiabatic derivative coupling elements for the coupled cluster singles and doubles model. The derivative coupling elements are evaluated in a  biorthonormal formulation in which the nuclear derivative acts on the right electronic state, where this state is biorthonormal with respect to the set of left states. This  
stands in contrast to earlier implementations based on normalized states and a gradient formula for the derivative coupling. As an illustration of the implementation, we determine a minimum energy conical intersection between the $n\pi^\ast$ and $\pi\pi^\ast$ states in the nucleobase thymine.

\end{abstract}

\maketitle

\section{Introduction}
The nuclear dynamics that follows photoexcitation typically involves non-adiabatic population transfer between several electronic states. For example, in the nucleobase thymine, photoexcitation to the bright $\pi \pi^\ast$ state is followed by rapid ($60$ fs) non-adiabatic population transfer to the dark $n \pi^\ast$ state.\cite{wolf2017probing} As is well known, the approximate description of the electronic structure can have a dramatic qualitative impact on the simulated nuclear dynamics, often complicating the task of correctly identifying the actual physics behind the processes observed in pump-probe experiments.\cite{domcke2011conical,curchod2018ab} A recent example is the ongoing debate about the dynamics that follows excitation to the bright $B_{3u}$ state in pyrazine.\cite{kanno2015ab,Horio2016,mignolet2018ultrafast,sun2020multi,scutelnic2021x} The ambiguities involved in interpreting time-resolved spectra illustrate the need for highly accurate description of the electronic structure. 

A number of electronic structure methods has a long history of being applied to treat non-adiabatic effects, including complete active space\cite{roos1980complete} (CAS) methods, density functional theory\cite{kohn1965self} (DFT), and algebraic diagrammatic
 construction\cite{schirmer1982beyond} (ADC). These methods are often complementary, where some are able to describe static correlation in the ground state and ground state intersections (CAS) while others better capture dynamical correlation but are unable to treat static correlation in the ground state as well as actual crossings with the ground state (DFT, ADC). In the latter category, there is still a need for a method that has systematically improvable accuracy that extends beyond a perturbative description of double excitations. 

 Coupled cluster theory is now well-established as the method of choice whenever this level of accuracy is required and the ground state is accurately described by a single determinant. However, initial progress towards its use in nonadiabatic dynamics simulations was slowed down with the realization\cite{hattig2005structure, kohn2007can} that the method produces non-physical results at electronic degeneracies when the  states that cross span the same symmetry. Later work by the present authors and collaborators showed that these artifacts were caused by the loss of electronic state orthogonality (matrix defects)\cite{kjonstad2017crossing} and that they could be fully removed by enforcing orthogonality relations between the electronic states.\cite{kjonstad2017resolving,kjonstad2019orbital} Our current understanding is that coupled cluster methods are able to describe conical intersections when the states span different symmetries but corrections\cite{kohn2007can,kjonstad2017resolving,kjonstad2019orbital} are required when the states span the same symmetry. However, these conclusions are based on studies of the potential energy surfaces and not from  considerations of the predicted physics. It  still remains an open question to what extent the artifacts at same-symmetry intersections negatively affect the predicted  dynamics  in trajectory-based simulation methods like surface hopping\cite{Tully1990} and \emph{ab initio} multiple spawning.\cite{Bennun2000} 

 Already in 1999 Christiansen\cite{Christiansen1999} derived expressions for the derivative coupling elements in coupled cluster theory, but the first implementation  was given later by Tajti and Szalay\cite{tajti2009analytic} at the singles and doubles level (CCSD). These authors did not, however, implement the expressions in Ref.~\citenum{Christiansen1999}. 
 Instead, the coupling was evaluated from the gradient of the two states as well as the gradient of a fictitious summed state; this summed-state approach was also used in a more recent  implementation of the CCSD coupling elements.\cite{Faraji2018}
 In addition, they proposed modifications to account for the fact that the coupled cluster states are not normalized, building on earlier work by Gauss and coworkers\cite{Gauss2006} who had found that normalization is important when evaluating the diagonal Born-Oppenheimer correction to the energy. The need for normalization in dynamics, which is not trivial to achieve, was later questioned by Shamasundar.\cite{shamasundar2018diagonal} In a recent publication, we confirmed this by showing that a biorthonormal formalism exists in which there is no dependence on the norm of the electronic states.\cite{Kjonstad2021}

 In the present work, we provide a derivation (which is equivalent to Ref.~\citenum{Christiansen1999}) and implementation, at the CCSD level of theory, of the derivative coupling between ground and excited states as well as between excited states. The derivation follows the Lagrangian approach for the derivative coupling proposed by Hohenstein in the context of CAS configuration interaction (CASCI),\cite{Hohenstein2016} while the present implementation builds on an efficient  implementation of analytical gradients, exploiting Cholesky decomposed electronic repulsion integrals, recently published by the authors and collaborators.\cite{schnackpetersen2022efficient} 

\section{Theory}

\subsection{Lagrangian}
The derivative coupling between states $i$ and $j$ is\cite{Christiansen1999,Kjonstad2021}
\begin{align}
    \mbf{F}_{ij} = \Dbraket{\psi_i^L}{\nabla \psi_j^R}, \quad i,j = 0,1,2,\ldots, \label{eq:nac}
\end{align}
where $L$ and $R$ signify that these are the left and right electronic states, and the gradient $\nabla$ is taken with respect to the coordinates of the atomic nuclei.

Analytical expressions for $\mbf{F}_{ij}$ may be derived by using the Lagrangian technique.
Here, we use the Lagrangian proposed by Hohenstein.\cite{Hohenstein2016} For the coupled cluster case, this Lagrangian can be expressed as\cite{Kjonstad2021}
\begin{align}
    \mathscr{L}_{ij} = \mathscr{O}_{ij} + \text{conditions}
\end{align}
where 
\begin{align}
    \mathscr{O}_{ij} = \Dbraket{\psi_i^L(\mbf{x}_0)}{\psi_j^R(\mbf{x})}.
\end{align}
Here we have made the dependence on the nuclear geometry explicit: $\mbf{x}_0$ is the geometry where the derivative is to be evaluated, while $\mbf{x}$ is allowed to vary. Upon differentiating $\mathscr{L}_{ij}$, the derivative operation $\nabla$ only acts on the ket vector. As a result, the derivative of $\mathscr{L}_{ij}$ at $\mbf{x}_0$ is identical to $\mbf{F}_{ij}$ at $\mbf{x}_0$.\cite{Hohenstein2016,Kjonstad2021}

The conditions in $\mathscr{L}_{ij}$ are those that are required to specify the right state $\psi_j^R$ for all values of $\mbf{x}$. These are: the Hartree-Fock equations, for specifying the orbitals; the amplitude equations, for specifying the ground state cluster amplitudes; and the excited state eigenvalue equations, for specifying the excited state amplitudes. Written out in detail, the Lagrangian reads
\begin{align}
\begin{split}
    \mathscr{L}_{ij} &= \mathscr{O}_{ij} + \sum_{\mu} \bar{\zeta}_\mu \Tbraket{\mu}{\bar{H}}{\hf}  \\
    &+ \sum_\mu \bar{\gamma}_\mu \bigl( \Tbraket{\mu}{[\bar{H}, R_j]}{\hf} - \omega_j R_\mu^j \bigr) \\
    &+ \bar{\xi} (1 - \Dbraket{L_j}{R_j}) + \sum_{ai} \bar{\kappa}_{ai} \mathscr{F}_{ai},
\end{split}
\end{align}
where we have suppressed the dependence on $\mbf{x}_0$ for notational convenience. 

This expression for $\mathscr{L}_{ij}$ introduces various quantities. The coupled cluster conditions are expressed in terms of the similarity-transformed Hamiltonian
\begin{align}
    \bar{H} = \exp(-T) \exp(\kappa) H \exp(-\kappa) \exp(T),
\end{align}
where we have introduced the orbital rotation operator
\begin{align}
    \kappa = \sum_{ai} \kappa_{ai} E_{ai}^-, \quad E_{ai}^- = E_{ai} - E_{ia},
\end{align}
as well as the cluster operator
\begin{align} 
T = \sum_\mu t_\mu \tau_\mu
\end{align}
The scalars $t_\mu$ are known as cluster amplitudes, and the $\tau_\mu$ denote excitation operators. The  $E_{ai}$ are singlet one-electron excitation operators and $E_{ia}$ are corresponding deexcitation operators. Here, $\kappa(\mbf{x}_0) = 0$ by assumption.

The electronic states are conveniently expressed as
\begin{align}
    \ket{\psi_k^R} &= \mathcal{R}_k \exp(T) \ket{\hf} \\
    \bra{\psi_k^L} &= \bra{\hf} \mathcal{L}_k \exp(-T)
\end{align}
where 
\begin{align}
    \mathcal{R}_k &= R_0^k + R_k = R_0^k + \sum_\mu R_\mu^k \tau_\mu \\
    \mathcal{L}_k &= L_0^k + L_k = L_0^k + \sum_\mu L_\mu^k \tau_\mu^\dagger.
\end{align}
We will also find it useful to write
\begin{align}
    \ket{\mathcal{R}_k} &= R^k_0 \ket{\hf} + \ket{R_k} \\ 
    \bra{\mathcal{L}_k} &= \bra{\hf}  L^k_0 + \bra{L_k}.
\end{align}
Furthermore, we have let
\begin{align}
    \omega_k = \Tbraket{L_k}{[\bar{H}, R_k]}{\hf}
\end{align}
and defined the Fock matrix as
\begin{align}
    \mathscr{F}_{pq} = h_{pq} + \sum_k (2 g_{pkkq} - g_{pqkk}).
\end{align}
Here, $h_{pq}$ and $g_{pqrs}$ are the one- and two-electron integrals of the Hamiltonian. Following the conventional notation, we let $p, q, r,$ and $s$ denote generic orbitals; $i, j, k,$ and $l$ denote occupied orbitals; $a,b,c,$ and $d$ denote virtual orbitals. Lagrangian multipliers are denoted with a bar ($\bar{\zeta}_\mu, \bar{\kappa}_{ai}$, $\bar{\xi}$).

The left-state quantities in $\mathscr{L}_{ij}$, that is, $\psi_i^L$ and $L_j$, are \emph{constants} that define $\mathscr{L}_{ij}$. They are evaluated at $\mbf{x}_0$. Thus, the Lagrangian's dependencies are understood as
\begin{align}
    \mathscr{L}_{ij} = \mathscr{L}_{ij}(\mbf{x}, \mbf{t}, \mbf{R}_j, \mbf{\kappa}, \bar{\mbf{\zeta}}, \bar{\xi}, \bar{\mbf{\kappa}}; \mbf{x}_0), 
\end{align}
where the semicolon denotes that $\mathscr{L}_{ij}$ depends only parametrically on $\mbf{x}_0$.

\subsection{Lagrangian stationarity conditions}
The derivative coupling becomes the partial derivative of $\mathscr{L}_{ij}$ when the Lagrangian is stationary with respect to all variables and multipliers that depend implicitly on $\mbf{x}$. We begin by considering stationarity for $\mbf{R}_j$:
\begin{align}
\begin{split}
    \frac{\partial \mathcal{L}_{ij}}{\partial R_\sigma^j} &= L^i_\sigma +  \sum_\mu \bar{\gamma}_\mu A_{\mu\sigma} \\
    &- \omega_j \bar{\gamma}_\sigma  - \sum_{\nu} L^j_\nu A_{\nu\sigma} \sum_\mu \bar{\gamma}_\mu R_\mu^j - \bar{\xi} L^j_\sigma = 0,
\end{split}
\end{align}
where
\begin{align}
    A_{\mu\nu} = \Tbraket{\mu}{[\bar{H}, \tau_\nu]}{\hf}.
\end{align}
Using vector notation, this condition reads
\begin{align}
    \mbf{0} = \mbf{L}_i^T + \bar{\mbf{\gamma}}^T (\mbf{A} - \omega_j) - (\omega_j \bar{\mbf{\gamma}}^T \mbf{R}_j + \bar{\xi})\mbf{L}_j^T
\end{align}
Clearly, with $\bar{\xi} = - \omega_j \bar{\mbf{\gamma}}^T \mbf{R}_j$, the last term in the equation vanishes, and we
 obtain stationarity provided
\begin{align}
   \bar{\mbf{\gamma}}^T = \frac{1}{\omega_j - \omega_i} \mbf{L}_i^T. 
\end{align}
We thus see that the excited state multipliers ($\bar{\xi}, \bar{\mbf{\gamma}}$) can be expressed in terms of the excited states ($\mbf{L}_i$, $\mbf{R}_j$) and the associated excitation energies ($\omega_i, \omega_j$).

Stationarity with respect to $\mbf{t}$ yields 
\begin{align}
    \mbf{0} = {^t}\mbf{\eta}^T + \bar{\mbf{\zeta}}^T \mbf{A} \label{eq:amplituderelax}
\end{align}
where
\begin{align}
    {^t}\eta_\sigma = \Tbraket{\mathcal{L}_i}{\tau_\sigma}{\mathcal{R}_j}  + (\mbf{\mathcal{F}}(\bar{\mbf{\gamma}})\mbf{R}_j)_\sigma,
\end{align}
with the well-known\cite{Koch1990} $F$-matrix defined as
\begin{align}
    \mathcal{F}(\bar{\mbf{\gamma}})_{\mu\nu} = \Tbraket{\bar{\gamma}}{[[\bar{H}, \tau_\mu],\tau_\nu]}{\hf}, \quad \bra{\bar{\gamma}} = \bra{\sigma} \bar{\gamma}_\sigma.
\end{align}
Similarly, stationarity with respect to $\mbf{\kappa}$ yields
\begin{align}
    \mbf{0} = {^\kappa}\mbf{\eta}^T + \bar{\mbf{\kappa}}^T \mbf{A}^\mathrm{HF}, \label{eq:orbitalrelax}
\end{align}
where
\begin{align}
\begin{split}
    {^\kappa}\eta_{ai} &= \Tbraket{\mathcal{L}_i}{E_{ai}^-}{\mathcal{R}_j} + \Tbraket{\bar{\zeta}}{[E_{ai}^-, \bar{H}]}{\hf} \\
    &+ \Tbraket{\bar{\gamma}}{[[E_{ai}^-,\bar{H}], R_j]}{\hf},
\end{split}
\end{align}
and where $\mbf{A}^\mathrm{HF}$ is the Hartree-Fock Hessian. The amplitude and orbital conditions, given by Eqs.~\eqref{eq:amplituderelax} and \eqref{eq:orbitalrelax}, are solved numerically for $\bar{\mbf{\zeta}}$ and $\bar{\mbf{\kappa}}$.

\subsection{Derivative coupling elements}
Once $\bar{\mbf{\zeta}}$ and $\bar{\mbf{\kappa}}$ are known, we can evaluate the coupling by taking the partial derivative of $\mathscr{L}_{ij}$ with respect to the nuclear components $\{ q \}$. This yields\cite{Christiansen1999, Kjonstad2021}
\begin{align}
    F_{ij}^q = \frac{\Tbraket{L_i}{[\bar{H}^q, R_j]}{\hf}}{\omega_j - \omega_i} + \Tbraket{\bar{\zeta}}{\bar{H}^q}{\hf} + \bar{\kappa}_{ai}  \mathscr{F}_{ai}^q, \label{eq:NAC_NC}
\end{align}
where
\begin{align}
    \bar{H}^q = \exp(-T) H^q \exp(T).
\end{align}
Here $H^q$ denotes the partial derivative of $H$ with respect to the $q$th nuclear coordinate, $x_q$. By expanding the commutator in Eq.~\eqref{eq:NAC_NC}, we obtain the equivalent expression
\begin{align}
    F_{ij}^q = \frac{\Tbraket{L_i}{\bar{H}^q}{R_j}}{\omega_j - \omega_i} + \Tbraket{\tilde{\zeta}}{\bar{H}^q}{\hf} + \bar{\kappa}_{ai}  \mathscr{F}_{ai}^q, \label{eq:NAC_NC_densities}
\end{align}
where 
\begin{align}
    \tilde{\mbf{\zeta}} = \bar{\mbf{\zeta}} - \mbf{J}, \quad J_\mu = \frac{\Tbraket{L_i}{R_j}{\mu}}{\omega_j - \omega_i} = \frac{j_{\mu}}{\omega_j - \omega_i}.
\end{align}
Clearly, $\mbf{F}_{ij}$ is the sum of an excited state gradient and a ground state gradient, plus an orbital relaxation term. The expression in Eq.~\eqref{eq:NAC_NC_densities} is convenient when invoking an existing molecular gradient code.

So far we have assumed that the right state ($\psi_j^R$) is an excited state. This raises the question of how to evaluate the coupling when $\psi_j^R$ is the ground state ($j = 0$). When this is the case, the excited state condition in $\mathscr{L}_{ij}$ can be removed. As a result, the $\mbf{t}$ stationarity simplifies to
\begin{align}
    \mbf{0} = \mbf{L}_i^T + \bar{\mbf{\zeta}}^T \mbf{A},
\end{align}
so that
\begin{align}
    \bar{\mbf{\zeta}}^T = - \frac{1}{\omega_i} \mbf{L}_i^T = \frac{1}{E_0 - E_i} \mbf{L}_i^T,
\end{align}
where $E_k$ denotes the electronic energy of the $k$th state. The orbital multiplier equation is also simplified by the removal of $\bra{\bar{\gamma}}$, but this equation must still be solved numerically. Once $\bar{\mbf{\kappa}}$ is known, we can evaluate $\mbf{F}_{i0}$ as
\begin{align}
    F_{i0}^q =  \frac{\Tbraket{L_i}{\bar{H}^q}{\hf}}{E_0 - E_i} + \bar{\kappa}_{ai}  \mathscr{F}_{ai}^q.
\end{align}

\subsection{Significance of orbital connections}

Hamiltonian derivatives are treated in the same way as for molecular energy gradients. That is, we take $H$ to be expressed, for all $\mbf{x}$, in a non-unique orthonormal MO (OMO) basis which is defined by an orbital connection.\cite{olsen1995orbital} Any orbital connection can be used, but the choice may actually affect the expression for the derivative coupling. In fact, as we will explain below, the formula in Eq.~\eqref{eq:NAC_NC} is only correct when we use the natural connection.\cite{olsen1995orbital,Christiansen1999, Kjonstad2021} 
For other connections, such as the widely-used symmetric connection, the partial derivative of $\mathscr{O}_{ij}$ is non-zero and must be added to the expression for $\mbf{F}_{ij}$.\cite{Hohenstein2016} 

 \label{sec:orbitalcon}

To show this, we express the derivative of $\mathscr{O}_{ij}$ in terms of the orbital connection. 
Given a connection matrix $\mbf{T}$, we define the OMOs as
\begin{align}
    \psi_p = \sum_q T_{pq} \phi_q,
\end{align}
where the unmodified MOs (UMOs) are given as
\begin{align}
    \phi_q = \sum_\alpha C_{\alpha q}(\mbf{x}_0) \chi_\alpha(\mbf{x}).
\end{align}
Here, $\{ C_{\alpha q} \}$ denotes MO coefficients, and $\{ \chi_\alpha \}$ denotes atomic orbitals. The UMOs are generally only orthonormal at $\mbf{x}_0$, that is,
\begin{align}
    S_{rs} = \Dbraket{\phi_r}{\phi_s} \neq \delta_{rs}, \quad \mbf{x} \neq \mbf{x}_0.
\end{align}
This is, of course, why an orbital connection is required in the first place; consistently evaluating the derivative is most easily done in a Fock space defined by an orbital basis that is orthonormal for all values of $\mbf{x}$.

\begin{figure*}[htb]
    \centering
    \includegraphics[width=\linewidth]{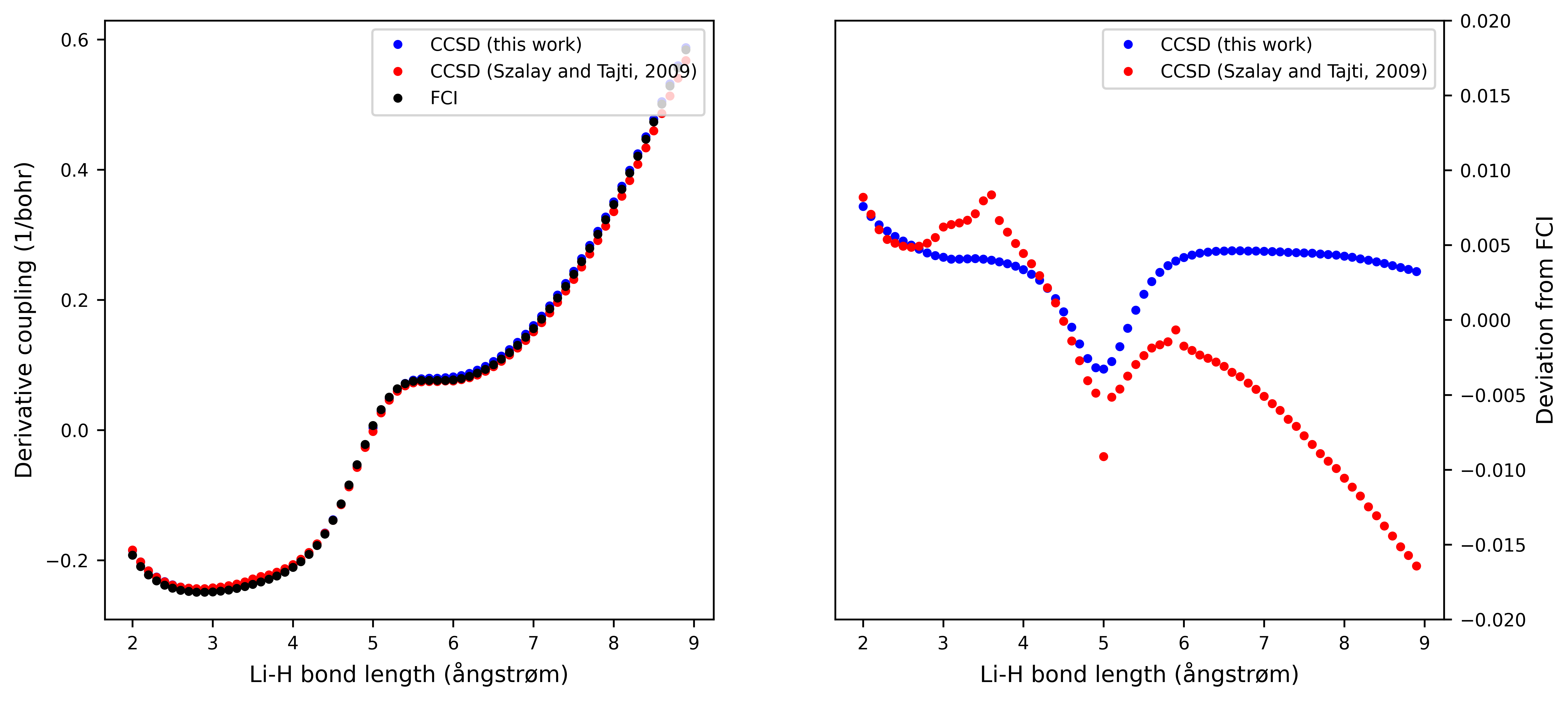}
    \caption{LiH/cc-pVQZ derivative coupling calculated for CCSD and FCI. For CCSD, we present couplings both for the direct evaluation of the coupling (this work) and for the summed-gradient values in Ref.~\citenum{tajti2009analytic}.}
    \label{fig:LiH_FCI_CC}
\end{figure*}

Now, the derivative of $\mathscr{O}_{ij}$ can be written\cite{olsen1995orbital}
\begin{align}
\begin{split}
    \mathscr{O}_{ij}^q = \frac{\partial \mathscr{O}_{ij}}{\partial x_q}\Big\vert_0
    = \sum_{rs} D_{rs}^{ij} Y_{rs}^q,
\end{split}
\end{align}
where $\mbf{D}^{ij}$ is the transition state density at $\mbf{x}_0$, and
\begin{align}
    Y_{rs}^q = \DBraket{\psi_r}{\frac{\partial \psi_s}{\partial x_q}}\Big\vert_0.
\end{align}
For the natural connection, we have, by construction,\cite{olsen1995orbital} 
\begin{align}
    Y_{rs}^q = 0,
\end{align}
and so we can conclude that\cite{olsen1995orbital,Christiansen1999}
\begin{align}
    \mathscr{O}_{ij}^q = 0.
\end{align}

Next, let us consider the symmetric connection. In this case, $\mbf{T} = \mbf{S}^{-1/2}$, which implies that
\begin{align}
    \frac{\partial T_{rs}}{\partial x_q} = -\frac{1}{2} \frac{\partial S_{rs}}{\partial x_q}\Big\vert_0 = -\frac{1}{2} (W_{rs}^q + W_{sr}^q),
\end{align}
where 
\begin{align}
    W_{rs}^q = \DBraket{\phi_r}{\frac{\partial \phi_s}{\partial x_q}}\Big\vert_0.
\end{align}
Consequently,
\begin{align}
    Y_{rs}^q = W_{rs}^q - \frac{1}{2} (W_{rs}^q + W_{sr}^q) = \frac{1}{2} ( W_{rs}^q - W_{sr}^q ),
\end{align}
and so
\begin{align}
\begin{split}
    \mathscr{O}_{ij}^q &= \sum_{rs} D_{rs}^{ij} \Bigl( \frac{1}{2} ( W_{rs}^q - W_{sr}^q ) \Bigr) \\
    &= \sum_{rs} \Bigl( \frac{1}{2} (D_{rs}^{ij} - D_{sr}^{ij}) \Bigr)W_{rs}^q.
\end{split}
\end{align}
For the symmetric connection, therefore, the derivative of $\mathscr{O}_{ij}$ is equal to the anti-symmetrized density matrix contracted with a ket-derivative of an overlap matrix.\cite{Hohenstein2016} This overlap derivative is evaluated as
\begin{align}
    W_{rs}^q = \sum_{\alpha \beta} C_{\alpha r} C_{\beta s} \DBraket{\chi_{\alpha}}{\frac{\partial \chi_\beta}{\partial x_q}\Big\vert_0}.
\end{align}

For the natural connection, $\mbf{W}^q$ is of course not needed for $\mathscr{O}_{ij}^q$ (which is zero). However, $\mbf{W}^q$ \emph{is} required for the reorthonormalization terms associated with the Hamiltonian. For the natural connection, the ket-derivative $\mbf{W}^q$ plays the same role that the braket-derivative $\mbf{S}^q$ does for the symmetric connection.\cite{olsen1995orbital} These reorthonormalization terms are the same for derivative couplings and molecular energy gradients, so we refer the reader to the literature for more details.\cite{schnackpetersen2022efficient}

\subsection{Relation to previous implementations} \label{sec:comparison}
In the literature, the derivative coupling has been implemented through a summed-state formula\cite{tajti2009analytic,Faraji2018} which is closely related to the one presented in this work. However, we have not been able to show that the two formulations are equivalent, except in the FCI limit. As we will see, our values for the coupling deviates to some extent from the values presented by Tajti and Szalay for the LiH molecule.\cite{tajti2009analytic}

\section{Implementation}
\subsection{Evaluation of the derivative coupling}
The derivative coupling has been implemented in a development version of the $e^T$ program.\cite{eT} The implementation builds on the recent implementation by Schnack-Petersen \emph{et al}.\cite{schnackpetersen2022efficient} for ground and excited state molecular gradients. Our implementation uses existing routines for molecular gradients and two-electron densities,\cite{schnackpetersen2022efficient} as well as several other quantities already implemented in the $e^T$ program,\cite{eT} such as the $F$-matrix ($\mbf{\mathcal{F}}(\bar{\mbf{\gamma}})$), the Hartree-Fock Hessian ($\mbf{A}_\mathrm{HF}$), and the second and third terms of ${^\kappa}\mbf{\eta}$. We apply central differences to obtain $\mbf{W}^q$ numerically, exploiting Libint 2\cite{valeev2020libint} to evaluate the AO overlap integrals. 

We have implemented the first term in the ${^t}\mbf{\eta}$ vector and in the ${^\kappa}\mbf{\eta}$ vector, that is, the terms that arise when differentiating $\mathscr{O}_{ij}$ with respect to $\mbf{t}$ and $\mbf{\kappa}$. In the case of CCSD, ${^t}\mbf{\eta}$ can be expressed as
\begin{align}
\begin{split}
    {^t\eta}_{ai}^1 &= L_{ai}^i R_0^j + j_{ai} \\
    &= L_{ai}^i R_0^j + \sum_{bj} L_{bj}^i R_{bjai}^j = D^{ij}_{ai}
\end{split}\\
    {^t\eta}_{aibj}^1 &= L_{aibj}^i R_0^j,
\end{align}
where $\mbf{D}^{ij}$ is the one-electron transition density. 
Finally:
\begin{align}
    {^\kappa\eta}_{ai}^1 = D^{ij}_{ai} - D^{ij}_{ia}.
\end{align}
We use an existing implementation to obtain the transition density $\mbf{D}^{ij}$.\cite{eT}

Finally, we have implemented the normalization factor $N_j^L$, since this allows us to validate our implementation by comparison to the exact limit. Programmable expressions for this quantity can be found elsewhere.\cite{tajti2009analytic}

\begin{figure*}[htb]
    \centering
    \includegraphics[width=\linewidth]{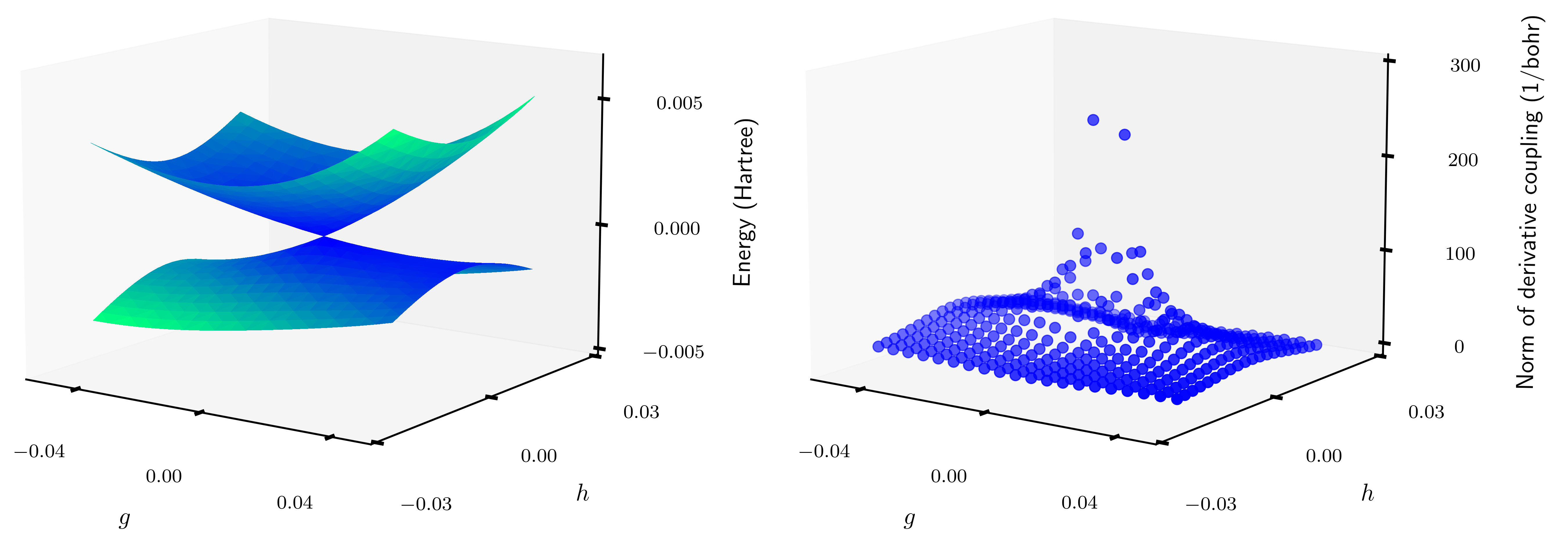}
    \caption{Branching plane for CCSD/aug-cc-pVDZ conical intersection in \ce{H2S} ($1 {^1} A_2 / 1 {^1} B_1$). We depict the  relative electronic energies (left) and the norm of the coupling vector (right).}
    \label{fig:branching_plane_h2s}
\end{figure*}

\begin{figure*}[htb]
    \centering
    \includegraphics[width=\linewidth]{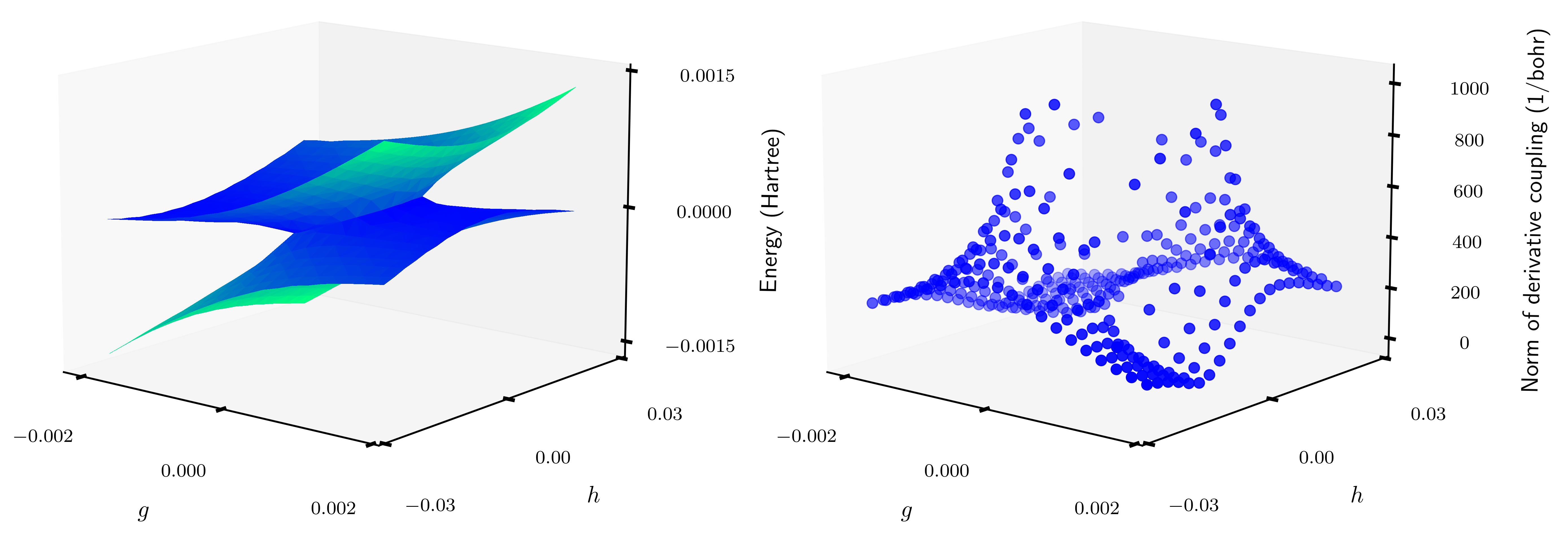}
    \caption{Branching plane for CCSD/aug-cc-pVDZ conical intersections in \ce{HOF} ($1 {^1}A'' / 2 {^1} A''$). We depict the real part of relative electronic energies (left) and the norm of the coupling vector (right).}
    \label{fig:branching_plane_hof}
\end{figure*}

\subsection{Optimization of minimum energy conical intersections} \label{sec:implementation_bfgs}
As numerical illustrations of the new implementation, we have applied Bearpark \emph{et al}.'s algorithm for determining minimum energy conical intersections (MECIs), where a gradient is constructed so that it is zero when two conditions are fulfilled: the energy difference vanishes and the energy gradient along the seam is zero.\cite{bearpark1994direct} In particular, we minimize the gradient
\begin{align}
    \mbf{G} = \mathcal{P} \nabla E_2 + 2 (E_2 - E_1) \frac{\mbf{g}}{\vert\vert \mbf{g} \vert \vert},
\end{align}
where 
\begin{align}
    \mbf{g} = \nabla (E_2 - E_1)
\end{align}
and  where $\mathcal{P}$ is the projection onto the complement of the $\mbf{g}$-$\mbf{h}$ plane. The $\mbf{h}$ vector is
\begin{align}
    \mbf{h} = (E_2 - E_1) \mbf{F}_{12}.
\end{align}
The gradient $\mbf{G}$ is used in combination with a Broyden-Fletcher-Goldfarb-Shanno (BFGS) solver already implemented in $e^T$ for geometry optimizations.\cite{schnackpetersen2022efficient}

\begin{figure*}[htb]
    \centering
    \includegraphics[width=\linewidth]{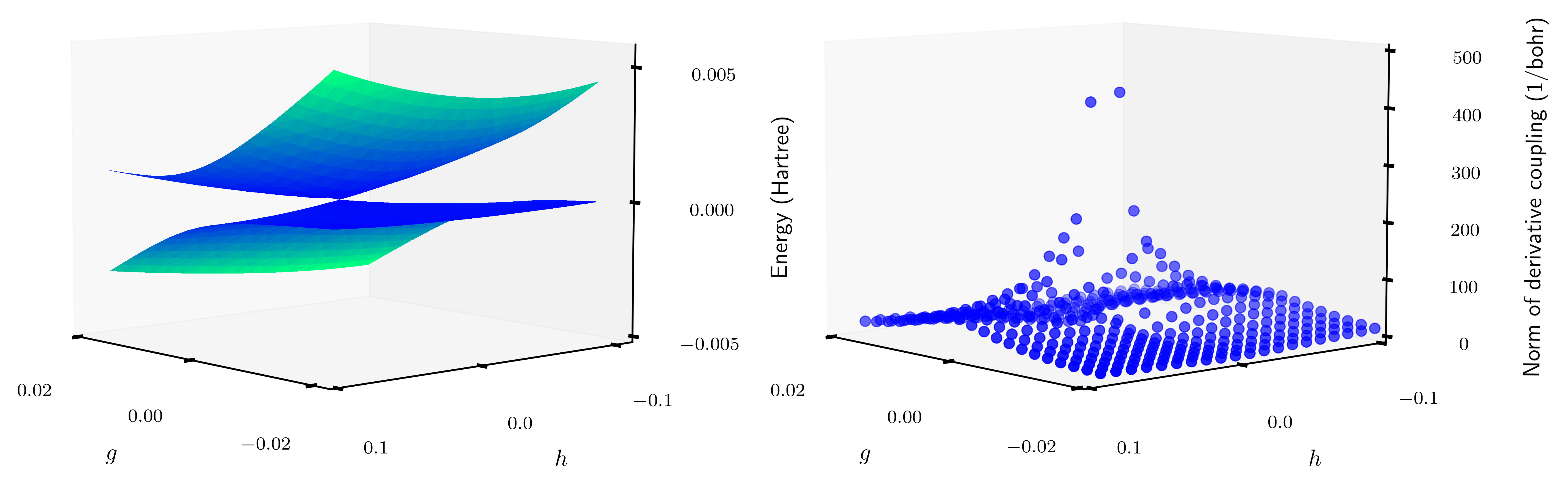}
    \caption{Branching plane for CCSD/cc-pVDZ C$_s$ minimum energy conical intersection in thymine ($n \pi^\ast$/$\pi \pi^\ast$). We depict the  relative electronic energies (left) and the norm of the coupling vector (right).}
\label{fig:branching_plane_thymine}
\end{figure*}

\section{Numerical examples}

\subsection{Comparison to earlier implementation: \ce{LiH}} 
In Figure \ref{fig:LiH_FCI_CC}, we show the derivative coupling element for the \ce{LiH} system as a function of the \ce{Li-H} bond distance, computed with three methods: CCSD using the direct formula (present work), CCSD using summed-state formula (numbers taken from Tajti and Szalay\cite{tajti2009analytic}), and the exact FCI derivative couplings (obtained with OpenMolcas\cite{aquilante2020modern}). All calculations are performed with the Dunning\cite{Dunning1989} basis cc-pVQZ. 

All three methods agree closely for all bond distances. However, there is a slight deviation between our results and that given in Ref.~\citenum{tajti2009analytic}, see Figure \ref{fig:LiH_FCI_CC} (right). This may be caused by both insufficient numerical convergence (as indicated by the uneven deviation from FCI) as well as differences in the analytical derivative couplings, as noted in Section \ref{sec:comparison}.

In order to ensure a consistent comparison to FCI, where states are normalized by default, we approximate the coupling from normalized coupled cluster states, averaging over the left and right coupling elements:
\begin{align}
\begin{split}
    \bar{\mbf{F}}_{ij}^\mathrm{norm} &= \frac{\mbf{F}_{ij}^\mathrm{norm} - \mbf{F}_{ji}^\mathrm{norm}}{2} \\
    &= \frac{\Dbraket{N_i^L \psi_i^L}{\nabla N_j^R \psi_j^R} - \Dbraket{N_j^L \psi_j^L}{\nabla N_i^R \psi_i^R}}{2}  \\
    &= \frac{N_i^L N_j^R \mbf{F}_{12} - N_j^L N_i^R \mbf{F}_{21}}{2} \\
    &\approx \frac{N_i^L (N_j^L)^{-1} \mbf{F}_{12} - N_j^L (N_i^L)^{-1} \mbf{F}_{21}}{2}
\end{split}
\end{align}
Recall that this normalization procedure is only required when we compare to methods with normalized states.

\subsection{Branching planes in three-atomic systems: \ce{SH2}, \ce{HOF}}
To provide some indication as to the behavior of the coupling in the vicinity of conical intersections, we have calculated branching planes for points of intersection in \ce{SH2} ($1 {^1} A_2 / 1 {^1} B_1$) and \ce{HOF} ($1 {^1}A'' / 2 {^1} A''$); see Figures \ref{fig:branching_plane_h2s} and \ref{fig:branching_plane_hof}, respectively. As expected, we find a divergence at the point of intersection in \ce{SH2} and no visible artifacts. This is consistent with the fact that this is an intersection between states spanning different symmetries.\cite{kjonstad2017crossing} The \ce{HOF} intersection, on the other hand, is defective because the states have the same symmetry. Note that the coupling still diverges as one approaches the defect.

\subsection{Minimum energy conical intersection: thymine}

Finally, we have applied the optimization algorithm described in Section \ref{sec:implementation_bfgs} to locate the $n \pi^\ast$/$\pi\pi^\ast$ minimum energy conical intersection in thymine, restricted to nuclear geometries with $C_s$ symmetry; see Figure \ref{fig:branching_plane_thymine}. In this calculation, we have used the cc-pVDZ basis. As for \ce{SH2}, this is a different-symmetry intersection and there is no sign of non-physical artifacts.

\section{Summary and outlook}
In this work we have presented an efficient implementation of derivative coupling elements that will enable us to perform large-scale simulations of nonadiabatic dynamics at the CCSD level of theory. Chemical systems of interest are now within the reach of CCSD dynamics using \emph{e.g.}~the multiple spawning framework;\cite{Bennun2000} for example, a single-point calculation on thymine with a cc-pVDZ basis, including gradients of the $n \pi^\ast$ and $\pi \pi^\ast$ states, as well as the coupling between them, can be performed in a matter of minutes on a modern CPU node (see Schnack-Petersen \emph{et al.}\cite{schnackpetersen2022efficient} for representative timings). 

We emphasize that for systems where the intersecting states span the same symmetry, the wavepacket may end up in regions that encompasses a defective  intersection. We then expect that corrections must be applied to the standard CC methods in order to extract meaningful results, though this will depend on the size of the defective intersection seam, which, in turn, depends on the truncation level. Work on extending the present implementation to the similarity constrained coupled cluster method (SCCSD), where such defects are completely eliminated,\cite{kjonstad2017resolving,kjonstad2019orbital} is in progress. Note that the Lagrangian approach makes such an extension straight-forward; we simply need to add the orthogonality condition to the Lagrangian and solve the resulting response equations.

The case that \emph{can} be treated with standard coupled cluster theory is that of intersections where the states span different symmetries (\emph{e.g.}~the $n \pi^\ast$ and $\pi \pi^\ast$ states in thymine). We may expect that such systems can be accurately described in dynamics simulations where coupled cluster theory provides the underlying electronic structure.  
This is the subject of a forthcoming article.

\begin{acknowledgments}
We thank David M.~G.~Williams for enlightening discussions. This work has received funding from the European Research Council (ERC) under the European Union’s Horizon 2020 Research and Innovation Programme (grant agreement No. 101020016). E.F.K. and H.K. both acknowledge funding from the Research Council of Norway through FRINATEK project 275506. We acknowledge computing resources through UNINETT Sigma2 -- the National Infrastructure for High Performance Computing and Data Storage in Norway, through project number NN2962k.

\end{acknowledgments}

\bibliography{apssamp}

\end{document}